\documentclass[referee,pdflatex,sn-nature]{sn-jnl}

\usepackage{graphicx}%
\usepackage{multirow}%
\usepackage{amsmath,amssymb,amsfonts}%
\usepackage{amsthm}%
\usepackage{mathrsfs}%
\usepackage[title]{appendix}%
\usepackage{xcolor}%
\usepackage{textcomp}%
\usepackage{manyfoot}%
\usepackage{booktabs}%
\usepackage{algorithm}%
\usepackage{algorithmicx}%
\usepackage{algpseudocode}%
\usepackage{listings}%

\usepackage{multibib}
\newcites{methods}{References}
\newcommand{\citemixed}[2]{[\citealp{#1},\citealpmethods{#2}]}

\theoremstyle{thmstyleone}%
\theoremstyle{thmstyletwo}%

\theoremstyle{thmstylethree}%

\begin{document}

\title[Article Title]{Joint dynamical-geophysical evidence for a limit cycle in the Galilean moons}


\author*[1,2]{\fnm{Amirhossein} \sur{Bagheri}}\email{abagheri@caltech.edu}

\author[3]{\fnm{Steven} \sur{D. Vance}}
\author[2]{\fnm{Jim} \sur{Fuller}}
\author[4]{\fnm{Brianna} \sur{Fernandez}}
\author[5]{\fnm{Mohit} \sur{Melwani Daswani}}
\author[6]{\fnm{Hauke} \sur{Hussmann}}
\author[3]{\fnm{Gregor} \sur{Steinbrügge}}
\author[7]{\fnm{Francis} \sur{Nimmo}}

\affil*[1]{\orgdiv{Division of Geological and Planetary Sciences}, \orgname{California Institute of Technology}, \orgaddress{\city{Pasadena}, \state{California}, \country{USA}}}
\affil*[2]{\orgdiv{Division of Physics, Mathematics and Astronomy}, \orgname{California Institute of Technology}, \orgaddress{\city{Pasadena}, \state{California}, \country{USA}}}
\affil[3]{\orgdiv{Jet Propulsion Laboratory}, \orgname{California Institute of Technology}, \orgaddress{\city{Pasadena}, \state{California}, \country{USA}}}
\affil[4]{\orgdiv{Department of Earth, Environment, and Planetary Sciences}, \orgname{Brown University}, \orgaddress{\city{Providence}, \state{Rhode Island}, \country{USA}}}
\affil[5]{\orgdiv{Earth‐Life Science Institute}, \orgname{Institute of Science Tokyo}, \orgaddress{\city{Tokyo},  \country{Japan}}}
\affil[6]{\orgdiv{Institute of Space Research}, \orgname{German Aerospace Center}, \orgaddress{\city{Berlin}, \country{Germany}}}
\affil[7]{\orgdiv{Department Earth and Planetary Sciences}, \orgname{University of California Santa Cruz}, \orgaddress{\city{Santa Cruz}, \state{California}, \country{USA}}}

\vspace{-2cm}

\abstract{Io, Europa, and Ganymede orbit in the Laplace mean-motion resonance, where their orbital and thermochemical evolution are strongly coupled \cite{ojakangas1986episodic, hussmann2004thermal, fischer1990thermal, tobie_etal25, veenstra2026good}.
Forced eccentricities sustain tidal dissipation, powering Io's volcanism and maintaining Europa's subsurface ocean \cite{lari2024nature,  dekleer_etal24, nimmo2026internal}.
Tidal heating depends on the moons' eccentricities, semimajor axes, and interior properties. Orbital evolution depends on how satellite dissipation affects the resonant dynamics \cite{tobie_etal25, nimmo2026internal}. 
This coupled evolution has not been treated in a self-consistent framework constrained by modern observations.
Here, we combine modern astrometric and geophysical measurements of migration rates, surface heat fluxes, tidal response, and moment of inertia \cite{veeder_etal94, lainey_etal09, casajus_etal21, park2025io, levin_etal26} with an orbital--thermochemical evolution model to constrain the long-term evolution and present state of the moons. 
We show that the joint observations select an ongoing late-time limit cycle, likely established $\sim$0.8--2.0~Gyr ago through feedback between thermal and orbital evolution.
The present-day state recurs within this oscillatory branch, whose cycles repeat every $\sim$95--150~Myr. 
Io presently experiences high dissipation and migrates inward, whereas Europa and Ganymede continue to migrate outward. 
Europa is predicted to undergo episodes of inward migration during the cycle.
During the cycles, Io's mean melt fraction varies substantially ($\sim$10--30\%), while Europa's ice-shell thickness varies by $\sim$3--15~km.
The predicted structures of Europa and Io will be assessed by future missions including Europa Clipper, JUICE, and IVO \cite{roberts2023exploring, zenk2025constraining, van2024geophysical, mcewen2014io}.}
\keywords{Galilean moons, Thermal--orbital evolution, Limit cycle, Tides}


\maketitle

The Laplace orbital resonance between Jupiter's large inner moons is likely primordial to the system, as evidenced by long-lived volcanism on Io \cite{dekleer_etal24} and by modern formation models of these satellites \cite{yap2025callisto}. 
Tidal dissipation couples the orbital and thermochemical evolution of the moons \cite{hussmann2004thermal, fischer1990thermal, veenstra2026good}.
This chain of interactions has controlled the differentiation of the moons, the overturning of their solid surfaces, and their volcanic histories \cite{sotin2002europa, chen2026temporal,mccarthy2016tidal,soderlund2020ice,bvehounkova2021tidally}. 
Tidal heating depends strongly on the orbital elements and rheological properties \cite{Peale1979IoTidalDissipation}:

\[ \dot E_{\rm tide} = \frac{21}{2} \frac{R^5 n^5}{G} e^2  (k_2/Q),\]

\noindent where $\dot E_{\rm tide}$ is the total tidal heating in the moon, $R$, $n$, and $e$ are the radius, mean motion, and orbital eccentricity, respectively, and $k_2/Q\equiv-\mathrm{Im}(k_2)$ is the dissipative component of the degree-2 tidal Love number of the moon, $k_2=\mathrm{Re}(k_2)+i\,\mathrm{Im}(k_2)$. 
This quantity links orbital evolution to internal structure which evolves as heat is produced and transported within the moon.

These moons orbit more slowly than Jupiter rotates and thereby receive a strong outward-driving torque from the planet \cite{nimmo2026internal}. 
Tidal dissipation within the moons tends to reduce the semimajor axes of the moons and competes with Jupiter's outward-pushing torque \cite{ojakangas1986episodic, boue2019tidal, bagheri_etal21}. 
Long-baseline astrometry shows that Io is currently migrating inward and Europa and Ganymede are migrating outward (Extended Data Table~\ref{edtable:constraints}) \cite{lainey_etal09}. 
Independent measurements constrain the interior structures of the moons. Io's tidal response and heat loss imply intense tidal heating and surface heat flux \cite{lainey_etal09,park2025io,mura2026synchronized,veeder_etal94,marchis_etal05}, while at Europa, Juno's passive microwave radiometry and Galileo radio science constrain the conductive thickness of the ice shell and the bulk moment of inertia \cite{casajus_etal21,levin_etal26}.
The latter has been used to propose partial differentiation and a cold evolution of Europa \cite{petricca2025partial,trinh2023slow}. 
These constraints have not been combined in a self-consistent thermal--orbital model; without them, previous studies allowed diverse plausible paths \cite{fischer1990thermal,hussmann2004thermal, bennacer2026divergent}.  

We model the coupled evolution by integrating the Laplace-resonance equations with self-consistent, layered thermochemical models of Io and Europa, while treating Ganymede as a point mass. 
As the interiors heat, partially melt, and differentiate, their evolving thermochemical states determine tidal heating, heat transport, and viscoelastic Love numbers, which feed back on orbital evolution.
We use Monte Carlo sampling over initial orbital elements, temperatures, and rheological parameters, including reference viscosities and Andrade parameters for rock \cite{bierson24, RenaudHenning18}.
We examine the present-day predictions against the astrometric--geophysical constraints (Extended Data Table~\ref{edtable:constraints}).
This framework connects the migration rates to interior evolution and tidal response. 
It accounts for melt-dependent rheology, silicate latent heat, and changes in shell properties, which together control the strength and timing of the moons' responses (Methods, Supplementary Sections~1,~2).

\subsection*{Observations select an ongoing limit-cycle}

\begin{figure}
    \centering
    \includegraphics[width=0.95\linewidth]{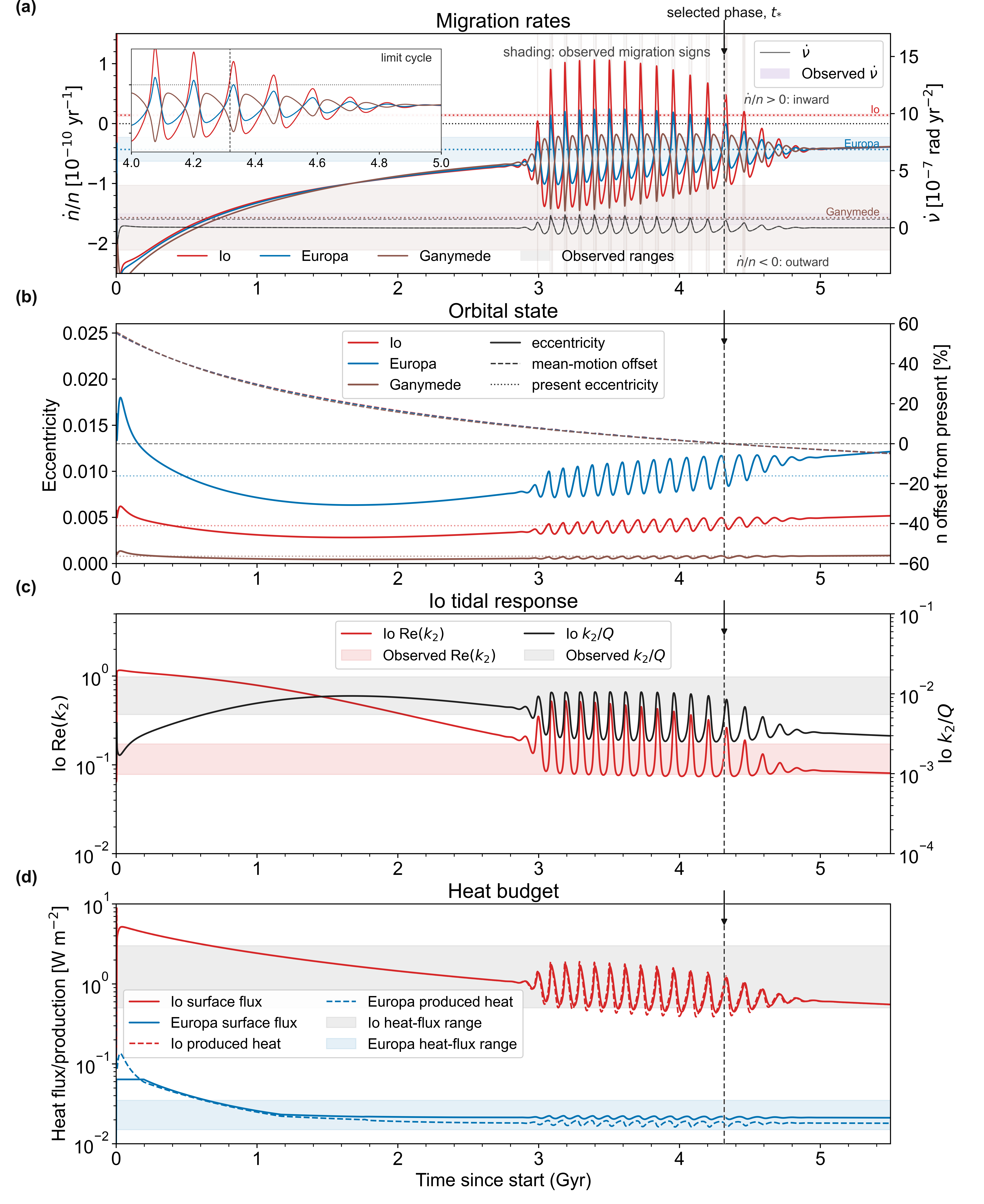}
    \caption{Representative orbital--thermochemical evolution history matching astrometric and geophysical constraints. 
    \textbf{(a)} Migration rates of Io, Europa and Ganymede, compared with the observed ranges \cite{lainey_etal09}. 
    Positive $\dot n/n$ corresponds to inward migration. 
    The right axis shows the resonance-drift diagnostics $\dot\nu=\dot n_1-2\dot n_2=\dot n_2-2\dot n_3$, compared with the measurement. 
    Dark shaded intervals mark phases in which Io migrates inward while Europa and Ganymede migrate outward; light shaded intervals mark phases in which both Io and Europa migrate inward. 
    The inset zooms into the late limit-cycle interval. \textbf{(b)} Evolution of the moons' eccentricities and mean-motion offsets from the present configuration.
    A relationship between the migration rates is derived in Supplementary~Section~4.
\textbf{(c)} Io tidal response, shown as ${\rm Re}(k_2)$ and $k_2/Q$, compared with the measured ranges \cite{park2025io}. 
\textbf{(d)} Internally produced heat and surface heat flux in Io and Europa. 
The total produced heat is divided by the surface area for comparison with heat loss.
The vertical dashed line marks the model phase $t_\ast\simeq4.3$~Gyr representing the present day (implying that the resonance formed in the first $\sim$0.2~Gyr).
}\label{fig1orbital}
\end{figure}

\begin{figure}
\centering
\includegraphics[width=1.\textwidth]{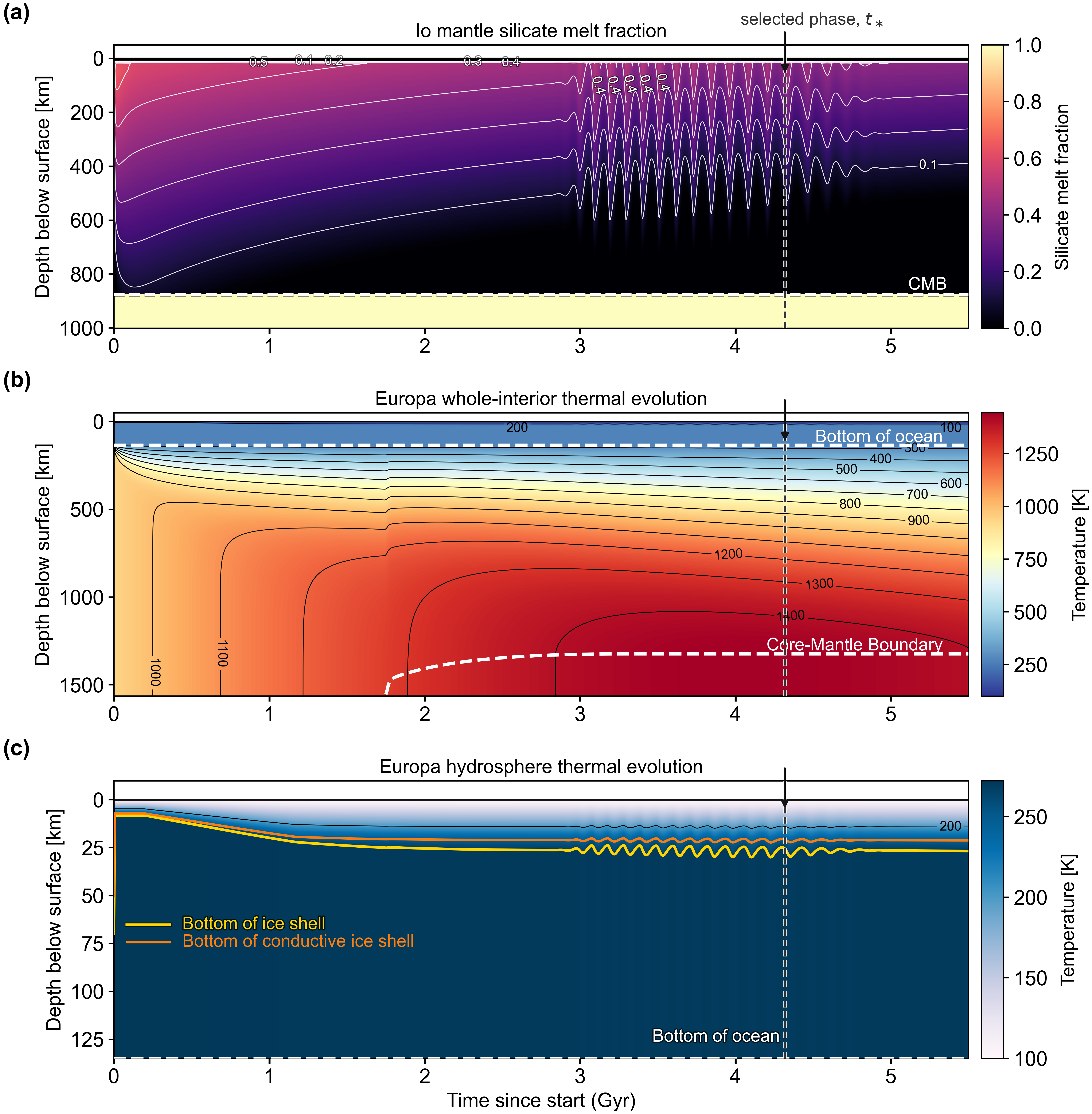}
\caption{Thermal and structural evolution of Io and Europa during the selected limit-cycle history. 
\textbf{(a)} Time--depth evolution of Io's silicate melt fraction. 
\textbf{(b)} Europa interior thermal evolution. 
The metallic core forms late when the mantle's maximum temperature reaches the Fe--FeS eutectic (Extended Data Fig.~\ref{phase_diagram}, Methods).
Tidal heating in Europa's mantle is small and core formation causes a small jump in mantle temperature. 
\textbf{(c)} Europa hydrosphere evolution, including conductive ice-shell thickness, total ice-shell thickness and ocean thickness. 
}\label{fig2thermal}
\end{figure}

The observed migration rates constrain the orbital-thermal state of the system. 
Fig.~\ref{fig1orbital} shows a representative orbital--thermochemical evolution matching the joint astrometric--geophysical constraints.
Upon initialization, the system evolves monotonically and then enters a late oscillatory state at $t\simeq3$~Gyr. 
The selected phase, $t_\ast\simeq4.3$~Gyr, reproduces the present-day migration rates, eccentricities, and mean motions (Fig.~\ref{fig1orbital}a, b) \cite{lari2024nature}.


The oscillating cycle is an outcome of the thermal-orbital coupling. 
This cycle begins when Io's upper mantle enters a temperature- and melt-sensitive rheological regime in which modest warming and melting reduce viscosity and rigidity, amplifying heat production.
Enhanced dissipation ($\propto e_{\rm Io}^2(k_2/Q)_{\rm Io}$) reduces the eccentricity that powers the heating. 
Diffusion and latent heat delay the thermal response of the mantle, causing the system to overshoot a steady balance.  
Io then cools and becomes less dissipative, allowing resonant forcing to restore its eccentricity and the sequence to repeat.
The system cycles between melt-rich, high-dissipation and colder, low-dissipation phases. 
As the mean thermal--orbital state drifts out of this feedback window, the oscillations decay and the resonance evolves smoothly.

Inward migration occurs during the high-dissipation phase of the cycle when the loss of orbital energy due to dissipation inside the satellites exceeds Jupiter's outward torque. 
The observed migration pattern requires that the orbital-energy loss, dominated by Io, exceeds the threshold required to drive Io's inward migration while remaining below the threshold of reversing Europa's migration (Extended Data Fig.~\ref{edfig:energy_balance}, Methods, Supplementary Sections~3,~4).



The selected phase also reproduces Io's measured $\mathrm{Re}(k_2)$ and $k_2/Q$, and the Io--Europa heat-flow constraints (Fig.~\ref{fig1orbital}b--d).
During oscillations, the moon-moon resonance offsets evolve jointly while  remaining dynamically coupled (Methods, Supplementary Section~4). 
In the representative history, the drift rate of the Laplace resonance, $\dot\nu\equiv\dot n_1-2\dot n_2\simeq\dot n_2-2\dot n_3$, matches the measurement \cite{lainey_etal09} near $t_\ast$ (Fig.~\ref{fig1orbital}a).

The near-equality between Io's heat production and surface heat loss \cite{veeder_etal94,park2025io} places the present state at the high-dissipation phase of the cycle.
This apparent thermal balance does not imply a steady thermal--orbital state, because inward migration cannot be sustained over long timescales. 
Instead, the joint constraints favor a transient phase during which dissipation drives inward migration while heat production remains close to the heat loss.

Fig.~\ref{fig2thermal} shows the interior evolution of the moons. 
In Io, the cycle is associated with partial melting in the upper mantle (Fig.~\ref{fig2thermal}a). 
At greater depths, the pressure-dependent solidus suppresses melting despite high temperatures (Supplementary Section~5). 
During hot phases, local melt fractions in the upper mantle exceed $\sim$40\%, whereas 
 they decrease to $\sim$30\% during colder phases. 
Dissipation is concentrated in regions that are hot enough to reduce viscosity  while retaining sufficient rigidity for efficient dissipation \cite{bagheri2025exploring}. %

Europa's forced eccentricity also varies in response to the cycle, resulting in changes in its tidal heating and interior properties (Methods).
In the representative model, Europa's ice shell  thickness oscillates by $\sim$5~km (Fig.~\ref{fig2thermal}b, c).
Melting and refreezing mostly occur near the base of the shell, where dissipation is largely concentrated (Supplementary Section~5).


\subsection*{Ensemble of evolutionary histories}

\begin{figure}
\centering
\includegraphics[width=1.0\textwidth]{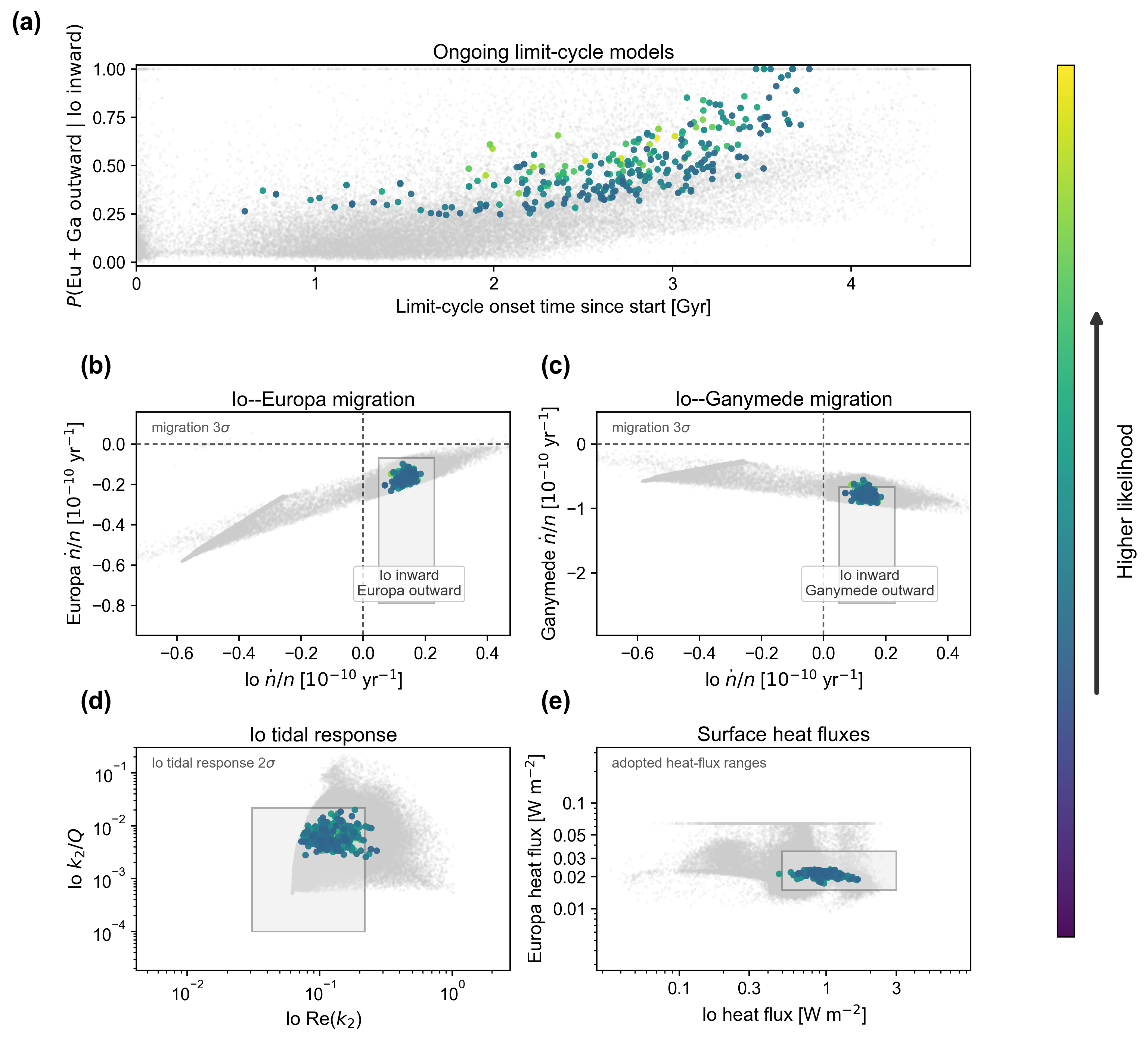}
\caption{Ensemble selection of coupled evolutionary histories.
\textbf{(a)} Onset time of the limit cycle and recurrence probability of the observed migration-sign state. 
The vertical axis shows the fraction of Io-inward intervals during which Europa and Ganymede continue migrating outward,
$P(\dot n_{\rm Eu}<0,\dot n_{\rm Ga}<0 \mid \dot n_{\rm Io}>0)$.
Grey points are simulations that develop a late cycle lasting to the present day.
Colored points show the high-scoring histories that satisfy the full astrometric and geophysical constraint set.
The colorbar represents the score of each of the high-ranking histories (Methods).
\textbf{(b,c)} Migration rates for Io--Europa and Io--Ganymede, compared with the measured range.
\textbf{(d)} Io tidal response, shown by ${\rm Re}(k_2)$ and $k_2/Q$, compared with the measured range.
\textbf{(e)} Surface heat fluxes of Io and Europa, compared with the observational ranges.
The joint astrometric--geophysical constraints favor an ongoing limit cycle; nearly 75\% of the highest-scoring histories place its onset between 0.8 and 2.0 Gyr before the present-day state.}
\label{ensemble}
\end{figure}

\begin{figure}
\centering
\includegraphics[width=1.\textwidth]{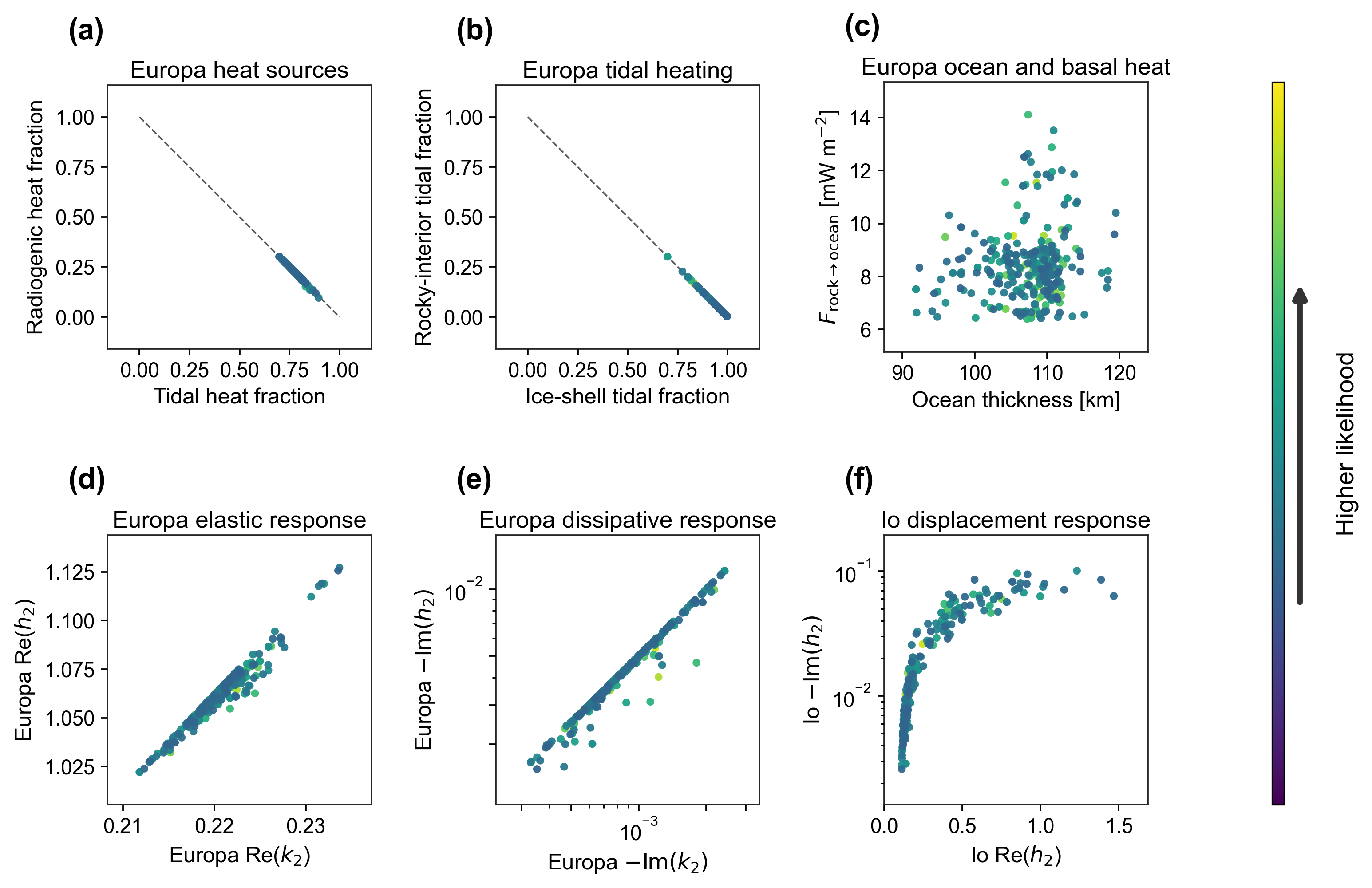}
\caption{Thermal diagnostics and predicted tidal observables of the high-score population.
\textbf{(a)} Europa internal heat-source partition at the selected phase, showing the relative contributions from tidal dissipation and radiogenic heating. Differentiation heating remains negligible.
\textbf{(b)} Partitioning of Europa's tidal dissipation between the ice shell and rocky mantle. 
\textbf{(c)} Europa ocean thickness and rock-to-ocean heat flux, showing the basal heat supply to the ocean. 
\textbf{(d)} Predicted elastic tidal response of Europa, shown by ${\rm Re}(k_2)$ and ${\rm Re}(h_2)$.
\textbf{(e)} Predicted dissipative tidal response of Europa, shown by the dissipative components $-\mathrm{Im}(k_2)$ and $-\mathrm{Im}(h_2)$. 
\textbf{(f)} Predicted displacement response of Io, showing $h_2$ and its dissipative component. 
The color scale matches Fig.~\ref{ensemble}.}
\label{fig4prediction}
\end{figure}

To assess whether cyclic behavior is generic, we analyzed an ensemble of $\sim$80,000 simulations in a Monte Carlo framework.
Each history was scored against the astrometric--geophysical constraints (Methods).
Of all histories, $31\%$ develop near-present oscillations. 
Fig.~\ref{ensemble} shows these models in grey and highlights the highest-scoring $1\%$ of the models that best satisfy all constraints. 

Non-oscillatory histories fail to reproduce Io's inward migration, whereas all the highest-scoring models develop near-present oscillatory behavior.
Fig.~\ref{ensemble}a quantifies the frequency of occurrence of the observed migrations. 
Ganymede migrates outward at all times, so the variation mainly reflects whether Europa has also crossed into inward migration. 
The late cycle begins a median of 1.6~Gyr before $t_\ast$, with 90\% of the high-score simulations placing its onset between 0.8 and 3.0~Gyr before $t_\ast$ and about 75\% between 0.8 and 2.0~Gyr.
In earlier-onset cycles, Europa more frequently migrates inward with Io, reducing the recurrence of the present-day configuration.

Io migrates inward for a median 22\% of the cycle.
The present configuration, $\dot n_{\rm Io}>0$, $\dot n_{\rm Eu}<0$ and $\dot n_{\rm Ga}<0$, occupies a median of 47\% of Io-inward intervals, so Europa's present outward migration represents a recurring subset of the Io-inward phase. 
Panels~b--e demonstrate that the high-score population fits all observations including the migration rates, Io tidal-response, and Io--Europa heat-flux ranges.



\subsection*{Europa's thermal state and differentiation}

Across the high-score models, Europa's late-cycle response is expressed mainly through its ice shell, with the mantle contributing $\lesssim15\%$ of the dissipation (Fig.~\ref{fig4prediction}a--c).
At $t_\ast$, the conductive and total ice-shell thicknesses are 19--23~km and 22--33~km thick, with late-cycle variations of 1--6~km and 3--15~km, respectively. 
Europa's heat budget is dominated by tidal dissipation, with radiogenic heating supplying the remainder and differentiation heating remaining negligible. 
The rock-to-ocean heat flux lies within 6.5--11.5~$\mathrm{mW\,m^{-2}}$, supplying a modest basal energy to the ocean.

Our selected histories predict that Europa's mantle may remain comparatively cold throughout the evolution, consistent with recent work \cite{trinh2023slow, petricca2025partial}. 
This inferred cold mantle depends on the adopted compositional model and the $C/(MR^2)$ constraint, which disfavor histories with strong mantle heating that would produce extensive metal segregation and formation of a large core (Discussion, Supplementary Section~6).
Maximum mantle temperatures span 1200--1650~K and are typically below silicate melting conditions, with $\lesssim$3\% of high-score models producing melt.
The mantle remains predominantly conductive and high-viscosity ($>10^{23}~\mathrm{Pa\,s}$), with limited convection in only 8\% of models. 
These constraints favor core radii of 90--550~km, with core formation beginning at $t\simeq2$~Gyr.

\subsection*{Io's melt-regulated evolution}
Io's observed $k_2/Q$ and heat production are controlled by a  melt-sensitive region that occupies roughly 10--25\% of the mantle, concentrated within the upper few hundred kilometers.
The median melt fraction of Io is 15\% at $t_\ast$, with 90\% of models within 8.5--25\%. 
During the cycle, the median mantle melt fraction (volume-averaged) is 19.5\%, and the lower- and higher-melt phases differ by a median melt fraction of 12\%.
During hot phases, maximum local melt fractions reach 46--52\%, and melt fractions above 10\% extend to depths of 525--685~km, favoring melt concentrated in the upper mantle 
(Supplementary Section~7).
 These results are consistent with the constraints from analyzing auroral spot oscillations \cite{roth2017constraints}.
The results also suggest that while significant melt fractions are produced in Io's history, a past magma ocean is not required.

\subsection*{Future measurements}

The simulations make testable predictions for observables targeted by future missions \cite{mazarico2023europa, rhoden2025diagnostic} (Fig.~\ref{fig4prediction}d--f). 
Most of the highest-scoring models predict Europa's elastic tidal response within $\mathrm{Re}(k_2)=0.21$--$0.23$ and $\mathrm{Re}(h_2)=1.04$--$1.09$, while its dissipative response is typically $-\mathrm{Im}(k_2)=3.0\times10^{-4}$--$2.5\times10^{-3}$ and $-\mathrm{Im}(h_2)=1.5\times10^{-3}$--$1.5\times10^{-2}$ (Fig.~\ref{fig4prediction}d,e).
For Io, most models predict a displacement Love number within $\mathrm{Re}(h_2)=0.12$--$1.2$ and $-\mathrm{Im}(h_2)=2.5\times10^{-3}$--$1.0\times10^{-1}$, reflecting the melt-sensitive anelastic interior states that regulate the cycle (Fig.~\ref{fig4prediction}f). 

In the oscillatory phase, Io undergoes large changes in tidal-response parameters and heat fluxes, with median variations of $9.0\times10^{-3}$ in $k_2/Q$ and 1.24~W\,m$^{-2}$ in surface heat flux, while Europa shows smaller but complementary shell-dominated variations. 
Interpreting Love numbers and heat fluxes together with gravity, rotational state, ephemerides and surface thermal emission can help identify the present orbital--thermal phase \cite{roberts2023exploring,van2024geophysical,christensen2024europa}. 

\subsection*{Discussion}
Io's observed inward migration provides key evidence that the Laplace--resonance system is not in a steady state.
This inward migration cannot be reproduced by the non-oscillatory histories.
Interpreting the observed migrations as a single isolated phase would require the present epoch to coincide with a short-lived transient phase. 
A more natural interpretation is that Io enters inward-migration intervals repeatedly during a cycle, with the present state sampling one such recurring phase.
Our favored histories provide this behavior while also satisfying Europa's observed outward migration.
The interior modules provide thermal--structural self-consistency, but the limit-cycle interpretation is driven primarily by the migration--geophysical observations.

The cyclic interpretation does not require the moons to leave the Laplace resonance.
They remain resonantly coupled, whereas the balance between Jupiter's forcing and satellite dissipation changes with the interior state. 
Given the current inward migration of Io and the heat budget, the measured drift rate $\dot\nu$ \cite{lainey_etal09} samples a phase in which the resonant offsets are near a maximum. 
As Io's tidal response weakens, Jupiter's torque and resonant coupling will drive those offsets toward smaller values over the subsequent $\sim$110~Myr.

Latent heat is a major control on the onset, termination and period of the limit cycle, and has not been explored in similar studies \cite{hussmann2004thermal, ojakangas1986episodic, fischer1990thermal}.
It modifies the thermal-orbital feedback by absorbing heat during melting and releasing heat during crystallization, increasing the effective heat capacity.
Without considering the latent heat, in the representative model, the onset of the cycle shifts from $t=3.0$ to $t=2.75$~Gyr, and it lasts for a longer time ($\sim$2~Gyr vs $\sim$1.5~Gyr), and the oscillation periods become shorter (Methods, Extended Data Fig.~\ref{edfig:latent_ablation}).

Europa's recent geology has been interpreted to indicate that the shell is thickening \cite{figueredo2004resurfacing,chen2026temporal}.
This observation is consistent with our inferred present state of the cycle, which lies in a declining-eccentricity phase, required by Io's inward migration, during which tidal heating decreases.
The inferred period of the cycle is $\sim$95--150~Myr, comparable to the $\sim$10$^8$~yr timescales proposed for cyclic variations of the resonance and of Europa's thermal state \cite{ojakangas1986episodic,fischer1990thermal,hussmann2004thermal, figueredo2004resurfacing}.
This timescale is also consistent with the mean geologic age from crater counts \cite{zahnle2008secondary}.


Our composition model assumes, as in recent studies \cite{trinh2023slow, petricca2025partial}, that Europa has a core of mostly iron with a small sulfur content. 
A larger, lower-density core enriched in light elements could also satisfy the $C/(MR^2)$ constraint and allow higher mantle temperatures, resulting in lower viscosities, higher dissipation, and possibly convection (Supplementary Section~6).
This non-uniqueness does not substantially affect the main thermal--orbital interpretation, because the cycle is controlled by Io's orbital-energy balance rather than the location of dissipation. 
Meanwhile, a thin dissipative basal layer in Europa's shell, as has also been suggested for Enceladus, Titan, and Ganymede, can explain the required heat \cite{tobie_etal05, kalousova2018two,petricca2025titan,bagheri2025exploring}.

The inferred dynamic changes to the shell thickness have implications for Europa's habitability.
The changing shell thickness may alter the ocean's composition. Chemical gradients generated by melting and refreezing could provide energy for life.
More important could be the cycling of material through the ice, which would bring solid deposits on the surface and the oxidants produced by radiolysis of surface ice to the ocean \cite{hesse2022downward}.
This efficient transport of materials through tidally heated ice could oxidize the ocean, a boon for habitability \cite{hand2007energy} if the ocean does not become overly acidic \cite{pasek2012acidification}. 
The possibility of a hotter, geophysically active mantle, produced either by a hot start or by tidal heating, counters recent studies arguing for an inactive mantle and less favorable conditions for habitability \cite{trinh2023slow, petricca2025partial, byrne2026little, green2025no}. 
Even in a cold mantle, fluids may percolate into a greater volume of rock in Europa's lower-gravity environment, and decreasing radiogenic heating could allow hydration reactions and radiolysis of water to produce appreciable amounts of H$_2$ \cite{vance2016geophysical, sherwoodlollar2014contribution}. 


Other heat-transport mechanisms in Io, including melt migration and heat-pipe volcanism, have been proposed \cite[e.g.,][]{lainey_etal09, o1981magma, sowell2026statistical}, but are not considered here for simplicity. 
Such mechanisms could modify the temperature, melting, and heat flux and may affect the timing, amplitude, and persistence of the cycle.
However, the underlying thermal-orbital feedback arises from Io's melt-sensitive dissipation, eccentricity damping, and resonant migration. 

Observables targeted by Europa Clipper, JUICE, and Io Volcano Observer through gravity, geodetic, thermal infrared, induction, and ephemeris measurements can test the inferred state and long-term evolution of the system \cite{roberts2023exploring,van2024geophysical,christensen2024europa,mcewen2014io}.
These results provide an observationally constrained basis for studying Io's volcanic history, Europa's seafloor volcanism, ocean--mantle exchange and habitability, geomorphology and tectonism, and the moons' dynamo histories. 
Our results show that combined dynamical--geophysical studies are key to understanding resonant systems that are common in the solar system and in exoplanetary systems.


\bibliography{sn-bibliography}

\clearpage
\section*{Methods}

\subsection*{Coupled thermal--orbital evolution}

We couple a Laplace-resonance orbital model to time-dependent, layered thermochemical models of Io and Europa. 
The interiors are represented by uniform layers whose temperature evolves with time, with melt fraction, viscosity, rigidity and density updated from the local thermal and compositional state.
Io is treated as a rocky body with melt-dependent rheology and a liquid core. 
Europa includes a rocky mantle, an ocean, and an ice shell that is divided into a conductive, largely elastic layer and a low-viscosity dissipative basal layer, and an evolving metallic core.
The layers are used for heat transport, melt evolution, density structure and the viscoelastic Love-number calculation, so changes in the thermochemical state modify the tidal response returned to the orbital model.

Europa's ice--ocean boundary temperature is updated from the local pressure as the shell thickness changes, and tidal heat generated in the ice shell is treated as an internal shell heat source. 
We partition tidal heating between the rocky mantle and ice shell by comparing the full layered Love-number response with a mantle-only response in which ice dissipation is suppressed and the difference gives the shell contribution.
The rocky interiors evolve by radial heat transport with tidal, radiogenic and differentiation heating. 
The thermal evolution is written as \citemethods{turcotte2014geodynamics}
\begin{equation}\label{eq:heat}
\rho c_{\rm eff}\frac{\partial T}{\partial t}=-\frac{1}{r^2}\frac{\partial}{\partial r}\left(r^2F\right)+ q_{\rm tide} +q_{\rm rad}+q_{\rm diff},
\end{equation}
where $\rho$ is density, $F$ is the radial heat flux, and $q_{\rm tide}$, $q_{\rm rad}$ and $q_{\rm diff}$ are the tidal, radiogenic, and differentiation heat sources. 
Radiogenic heat production follows the isotope inventory and abundances adopted by \cite{hussmann2004thermal}.
We use $c_{\rm eff}$ to include latent heat with an enthalpy formulation for phase change \citemethods{voller1987enthalpy}. 
For a layer with specific enthalpy $h$,
\begin{equation}
h(T,P)=c_pT+L\phi(T,P),
\end{equation}
$c_p$ is ordinary specific heat, $L$ is latent heat, $\phi$ is melt fraction, and $T$ and $P$ are local temperature and pressure. 
The thermal storage term can then be written in temperature form with
\begin{equation}
c_{\rm eff} =\left(\frac{\partial h}{\partial T}\right)_P = c_p + L\left(\frac{\partial \phi}{\partial T}\right)_P.
\end{equation}\label{eq:effheatcapac}
Outside the phase-change interval, $(\partial\phi/\partial T)_P=0$ and $c_{\rm eff}=c_p$. 
Across the solidus--liquidus interval, $c_{\rm eff}$ increases because part of the added or removed heat changes the melt fraction rather than the temperature. 
In the numerical model, this apparent heat capacity enters the thermal storage term of each shell. 
The derivative $\left(\frac{\partial \phi}{\partial T}\right)_P$ is computed from the lever-rule melt fraction described in the next subsection.

In the numerical model, equation~\ref{eq:heat} is applied shell by shell as
\begin{equation}
\rho_j c_{{\rm eff},j} V_j \frac{dT_j}{dt} = A_{j-1/2}F_{j-1/2} - A_{j+1/2}F_{j+1/2} + Q_{{\rm tide},j} + Q_{{\rm rad},j} + Q_{{\rm diff},j},
\end{equation}
where $V_j$ is the shell volume, $A_{j\pm1/2}$ are the areas of its inner and outer boundaries, and $F_{j\pm1/2}$ are the corresponding outward heat fluxes.
The terms $Q_{\rm tide,j}$, $Q_{\rm rad,j}$, and $Q_{\rm diff,j}$ are the integrated tidal, radiogenic, and differentiation heat sources within the shell.
The source terms are assigned to the corresponding thermal reservoirs or layers, allowing the partitioning of heating among the rocky mantle and ice shell to affect the subsequent thermal and rheological evolution.

Heat transport includes conduction and parameterized convection. 
The conductive flux is
\begin{equation}
F_{\rm cond}=-k\frac{\partial T}{\partial r}.
\end{equation}
where $k$ is the thermal conductivity. 
For potentially convecting layers, we compute a Rayleigh number \citemixed{hussmann2004thermal}{schubert_etal04,bagheri_etal22}
\begin{equation}
Ra=\frac{\rho g\alpha_T \Delta T D^3}{\kappa \eta_{\rm eff}},
\end{equation}
where $\alpha_T$ is thermal expansivity, $\Delta T$ is the temperature contrast across a layer of thickness $D$, $\kappa$ is the thermal diffusivity, and $\eta_{\rm eff}$ is the representative layer viscosity.
Convective heat transport is represented with a Rayleigh--Nusselt scaling \cite[e.g.,][]{hussmann2004thermal,chen2026temporal}, 
\begin{equation}
Nu=\max\left[1,\left(\frac{Ra}{Ra_c}\right)^{\beta_{\rm Nu}}\right],
\end{equation}
with $Ra_c=10^3$ and $\beta_{\rm Nu}=0.3$ \cite{hussmann2004thermal}. 
The radial heat flux is then written as $F=NuF_{\rm cond},$ so that $Nu=1$ recovers purely conductive heat transport and $Nu>1$ represents the enhancement due to convection. 
Europa's ice shell is evolved from the heat balance across its conductive and viscoelastic layers, including tidal heating generated within the shell.

At each timestep, the orbital state determines the tidal forcing, the layered interiors update their thermal and rheological structures, and the resulting Love numbers feed back into the next orbital step. 
The tidal calculation provides the produced heat and, for Europa, partitions it between the rocky mantle and the ice shell. 
The evolving layer densities are also used to compute the normalized moment of inertia of Europa and Io, allowing the interior structure that sets the tidal response and heat budget to be tested against the gravity constraint. 

For the orbital evolution, we use the reduced Laplace-resonance migration equations of \cite{fischer1990thermal,hussmann2004thermal}. 
We write the equations in terms of the mean motions $n_i$, where $i=1,2,3$ denotes Io, Europa and Ganymede. 
Since $n_i\propto a_i^{-3/2}$, positive $\dot n_i$ corresponds to inward migration and negative $\dot n_i$ corresponds to outward migration.
The migration rates are written in terms of the eccentricities and dissipative tidal responses of Jupiter, Io and Europa as \cite{hussmann2004thermal}
\begin{align}
\dot n_1 &= n_1^{16/3} \left[
X_1K_J + X_2K_1e_1^2 + X_3K_2E_X \right],
\\
\dot n_2 &= \frac{n_1^{16/3}}{2} \left[ (X_1-Y_1)K_J + (X_2-Y_2)K_1e_1^2 + X_3K_2\Delta E_{XY} \right],
\\
\dot n_3 &= \frac{n_1^{16/3}}{4} \left[ (X_1-3Y_1)K_J + (X_2-3Y_2)K_1e_1^2 + X_3K_2\Delta E_{X-3Y} \right].
\end{align}
Here, $K_J$, $K_1$ and $K_2$ are the tidal-dissipation coefficients for Jupiter, Io, and Europa. 
Their dimensional factors and coefficients $X_i$ and $Y_i$ follow the reduced Laplace-resonance formulation of \cite{fischer1990thermal,hussmann2004thermal}.
They contain the resonant coupling and dimensional factors and carry the signs of the orbital response. 
For Io and Europa, the time-dependent part of $K_1$ and $K_2$ is set by the evolving dissipative Love-number component, $(k_2/Q)_i=-\mathrm{Im}(k_{2,i})$. 
The eccentricity combinations are
\begin{align}
E_X &= X_4e_2^2+X_5e_{21}^2+X_6e_{23}^2,\\
\Delta E_{XY} &= (X_4-Y_4)e_2^2+(X_5-Y_5)e_{21}^2+(X_6-Y_6)e_{23}^2,\\
\Delta E_{X-3Y} &= (X_4-3Y_4)e_2^2+(X_5-3Y_5)e_{21}^2+(X_6-3Y_6)e_{23}^2.
\end{align}
Europa's eccentricity is written as $e_2=e_{21}+e_{23}$, where $e_{21}$ and $e_{23}$ are the components forced by Io and Ganymede. 
In this formulation, the Jovian term supplies angular momentum and orbital energy to the resonant chain, while dissipation inside Io and Europa removes orbital energy and produces heat within the satellites. 
Ganymede is not assigned a separate local dissipation term in the migration equations.
Its migration follows from the redistribution of angular momentum and orbital energy through the Laplace chain.

The drift of the resonant chain is computed as \cite{lainey_etal09}
\begin{equation}
\dot\nu=\dot n_1-2\dot n_2=\dot n_2-2\dot n_3 .
\end{equation}
This quantity is compared directly with the astrometric constraint. 
The model assumes that the Laplace resonance is maintained and the perturbations caused by the thermal-orbital feedback are small. 
This assumption is supported by the observed migration rates \cite{lainey_etal09}, and is reasonable given the large tidal forcing of Jupiter which maintains the moons in the resonance. 
It is also supported by the inference of sustained volcanism in Io \cite{dekleer_etal24}.

The relative amplitudes and phases of the late-time migration-rate
oscillations are also consistent with the approximate resonant relations
derived in Supplementary Section~4. 
For the representative model, linear comparisons of the oscillatory components give coefficients of $0.498$ for Europa and $-0.517$ for Ganymede relative to Io, close to the approximate values of $+0.50$ and $-0.50$ predicted analytically.

Thus, the orbital part follows the established reduced resonant framework, while the present calculation closes the feedback with evolving layered heat transport, melt-dependent rheology, latent heat and self-consistent tidal responses rather than prescribed satellite tidal properties.

The orbit and interiors are integrated as a coupled ODE system using an adaptive Runge--Kutta method. 
The state vector includes the resonant orbital variables, scalar thermal and structural variables such as Europa's ice-shell thicknesses and core radius, and the layered finite-volume mantle temperatures of Io and Europa. 
At each ODE evaluation, the state variables are mapped to layered density, viscosity and rigidity profiles, which are converted to complex shear moduli using the Andrade rheology \citemethods{bagheri_etal19, castillorogez_etal11, renaud_etal21, petricca_etal24}. 
The resulting profiles are passed to a degree-2 matrix-propagation Love-number solver for a spherically symmetric, self-gravitating viscoelastic body. 
The solver returns the complex Love numbers $k_2$ and $h_2$, and the positive dissipative component $k_2/Q=-\mathrm{Im}(k_2)$ is used in both the tidal-heating and orbital-migration equations. 
Further numerical details, including the state-vector size and Love-number layer counts, are given in Supplementary Section~1.

\subsection*{Europa metal--silicate differentiation and core formation}

Europa's metal--silicate differentiation is treated as part of the evolving layered interior.
Building from previous work \cite{petricca2025partial}, the metallic component is assumed to be an Fe--FeS alloy derived from an adopted CM-chondritic bulk composition \citemethods{lodders_planetary_1998} geochemically consistent with the silicate portion. 
The available Fe--S budget is fixed by this bulk composition, and the evolving metallic core radius $r_c$ is included as an additional state variable. 
Io is initialized with a metallic core and silicate mantle (tidal heating and accretion heating result in fast differentiation), whereas Europa's core is allowed to grow during the integration.

The metallic Fe–S melting interval was computed from pressure-dependent fits to the solidus and liquidus extracted from the CM-composition phase diagram calculated with Gibbs free energy minimization code Perple\_X \citemethods{connolly_geodynamic_2009} and appropriate thermodynamic models \citemethods{saxena_thermodynamics_2015} shown in Extended Data Fig.~\ref{phase_diagram}. 
We use
\begin{align}
T_{\rm sol}^{\rm metal}(P) &= 1268.6 + 14.95P - 4.23P^2,\\
T_{\rm liq}^{\rm metal}(P) &= 1385.36 + 123.6P - 7.14P^2.
\end{align}
where $P$ is in GPa and temperatures are in K. 
We use the metallic solidus and liquidus in a lever rule to compute the local metallic melt fraction used to grow the core:
\begin{equation}
\phi_{\rm metal}(r,t)= \begin{cases} 0, & T(r,t)\le T_{\rm sol}^{\rm metal}(P(r)),\\[3pt] \dfrac{T(r,t)-T_{\rm sol}^{\rm metal}(P(r))} {T_{\rm liq}^{\rm metal}(P(r))-T_{\rm sol}^{\rm metal}(P(r))}, & T_{\rm sol}^{\rm metal}(P(r))<T(r,t)<T_{\rm liq}^{\rm metal}(P(r)),\\[8pt] 1, & T(r,t)\ge T_{\rm liq}^{\rm metal}(P(r)).
\end{cases}
\end{equation}
For diagnostic quantities such as the mean metallic melt fraction, we compute the volume-weighted average,
\begin{equation}
\bar{\phi}_{\rm metal}(t)=\frac{\int_{r_c}^{R_{\rm rock}}\phi_{\rm metal}(r,t)r^2\,dr}{\int_{r_c}^{R_{\rm rock}}r^2\,dr}.
\end{equation}

Metal segregation is activated once the mantle-side core--mantle-boundary temperature reaches the local Fe--FeS solidus.
The inventory-limited target core mass is set by the available metal fraction, $\gamma_{\rm Fe}$, between a seed core mass and the maximum core mass allowed by the adopted bulk composition. 
The corresponding target radius, $r_{c,\ast}$, is computed for the adopted core density. 
The core radius then relaxes toward this target radius on a short segregation timescale $\tau_c$,
\begin{equation}
\frac{dr_c}{dt}=\frac{r_{c,\ast}-r_c}{\tau_c}.
\end{equation}

As the core grows, the total rocky mass is conserved by updating the density of the remaining silicate mantle,
\begin{equation}
\rho_m(t)=\frac{M_{\rm rock}-\rho_c(4\pi r_c^3/3)}
{(4\pi/3)(R_{\rm rock}^3-r_c^3)}.
\end{equation}
The evolving core radius and mantle density are used in the Love-number
and moment-of-inertia calculations.
The metallic core state is diagnosed from
the Fe--FeS melt fraction evaluated at the mantle-side core--mantle-boundary
temperature. 

In the integrations, the metallic core is thermally passive; its melt state is obtained from the mantle-side core--mantle-boundary temperature. 
Gravitational energy released by metal segregation is deposited into Europa's rocky mantle as differentiation heating,
\begin{equation}
Q_{\rm diff}=\epsilon_{\rm diff}\dot M_c\Delta\Phi,
\end{equation}
where $\Delta\Phi$ is the gravitational potential drop of the segregating metal:
\begin{equation}
\Delta\Phi = \int_{r_c}^{r_{\rm src}} \frac{Gm(r)}{r^2}\,dr, \qquad r_{\rm src}=r_c+\frac{2}{3}(R_{\rm rock}-r_c).
\end{equation}
Here $m(r)$ is the enclosed mass and $r_{\rm src}$ approximates the source radius of segregating metal in the rocky mantle, and $\epsilon_{\rm diff}$ is a factor determining the partitioning of the energy between the mantle and the core. 
We find this energy source to be unimportant compared to radiogenic heating and tidal heating (see Supplementary Section~5) and for simplicity assume $\epsilon_{\rm diff}=1$, so all the energy released is deposited in the mantle.

\subsection*{Silicate melting, melt-dependent rheology and tidal response}

Silicate melting is computed locally in each rocky layer from the layer temperature and pressure. 
We use pressure-dependent silicate solidus and liquidus curves \citemethods{takahashi1990speculations,vlaar1994cooling},
\begin{align}
T_{\rm sol}^{\rm sil}(P)&= 1409.15 + 134.2P - 6.581P^2 + 0.1054P^3,\\
T_{\rm liq}^{\rm sil}(P)&= 2035.15 + 57.46P - 3.487P^2 + 0.0769P^3,
\end{align}
where $P$ is in GPa and temperatures are in K. 
The local silicate melt fraction is evaluated with a lever rule analogous to that used for metallic melting. 
The pressure dependence allows computing partial melt as a function of depth. 
The local melt fraction enters both the thermal and tidal parts of the calculation. 
In the energy equation, the lever-rule derivative gives
\begin{equation}
\frac{\partial \phi_{\rm sil}}{\partial T}= \left(T_{\rm liq}^{\rm sil}-T_{\rm sol}^{\rm sil}\right)^{-1}
\end{equation}
inside the partial-melt interval and zero outside it. 
This equation is also used to compute the effective heat capacity $c_{\rm eff}$. 
In the tidal calculation, $\phi_{\rm sil}$ modifies the local viscosity and rigidity.
Below the solidus, the viscosity follows a subsolidus temperature-dependent branch \cite{hussmann2004thermal},
\begin{equation}\label{eq:Arrhenian}
\eta_{\rm sub}(T,P)=\eta_0\exp\left[\frac{E_A}{\hat R}\left(\frac{1}{T}-\frac{1}{T_{\rm sol}^{\rm sil}(P)}\right)\right],
\end{equation}
where $\eta_0$ is the reference viscosity at the local solidus, $\hat R=8.3143~{\rm J\,mol^{-1}\,K^{-1}}$ is the gas constant, and $E_A$ is the activation energy, taken to be 360 and 370~kJ~mol$^{-1}$ for Io and Europa, respectively \cite{hussmann2004thermal}. 
Below the solidus, the rock shear modulus is held fixed at $\mu_{\rm sol}=50~\mathrm{GPa}$.

We do not extrapolate equation~\ref{eq:Arrhenian} into the supersolidus regime. 
Instead, the local solidus--liquidus interval defines the thermodynamic silicate melt fraction
used for latent heat, melt diagnostics, and empirical weakening of the effective viscoelastic aggregate \citemixed{tobie_etal25}{costa2005viscosity,kervazo2021solid}. 
For $X$, with $X=\eta$ or $\mu$, the dimensionless weakening shape is defined as

\begin{equation}
C_X(\phi_{\rm sil})=\frac{1+\theta_X^{\delta_X}}{\left[1-F_X(\theta_X)\right]^{B\Phi_X}}, \qquad \theta_X=\frac{1-\phi_{\rm sil}}{\Phi_X},
\end{equation}
where
\begin{equation}
F_X(\theta_X)=(1-\xi_X)\operatorname{erf}\left[\frac{\sqrt{\pi}}{2(1-\xi_X)}\theta_X \left(1+\theta_X^{\gamma_X}\right)\right].
\end{equation}
Here $B=2.5$, and $\delta_X$, $\xi_X$, $\gamma_X$ and $\Phi_X$ control the sharpness and location of the weakening transition \citemethods{costa2005viscosity,kervazo2021solid}. 
We apply this formulation to the local supersolidus melt fraction, $\phi_{\rm sil}$, to map the effective viscosity and rigidity from their solidus values, $\eta_0$ and $\mu_{\rm sol}$, toward the bulk rheology of partially molten silicate material in the interior model.
This rheology captures the sharp changes in viscosity and rigidity inferred for partially molten rocky materials in experimental studies \citemethods{costa2005viscosity,costa2009model}. 

The radial viscosity and rigidity profiles are converted into complex shear moduli for each layer. 
We use Andrade anelasticity where the complex compliance is
\begin{equation}
J^\ast = \frac{1}{\mu}-\frac{i}{\omega\eta} +\beta_A\Gamma(1+\alpha_A) (i\omega_A)^{-\alpha_A}, \qquad \mu^\ast=\frac{1}{J^\ast}.
\end{equation}
Here, $\eta$ and $\mu$ are the local viscosity and rigidity, $\omega$ is the tidal forcing frequency, and $\alpha_A$ and $\beta_A$ are Andrade parameters \citemethods{Andrade62,bagheri_etal19,bierson2016test}. 
The pseudo-frequency master variable is
\begin{equation}
\omega_A = \omega \exp \left[ \frac{E_A}{\hat R} \left(\frac{1}{T} - \frac{1}{T_{\rm ref}} \right)\right].
\end{equation}
where $E_A$ is the activation energy for the temperature shift, $T_{\rm ref}$ is the reference temperature \citemethods{JacksonFaul10} taken to be the local silicate solidus temperature.
The resulting radial profiles of $\mu^\ast$, density and layer radius are passed to the layered viscoelastic Love-number solver.
The solver returns the complex degree-2 Love numbers $k_2$ and $h_2$, and $k_2/Q$ is used in both the tidal-heating and migration equations.

\subsection*{Monte Carlo sampling and scoring}

We explored the coupled model using Monte Carlo sampling of possible initial conditions and material parameters \citemethods{robert2004montecarlo}. 
The sampled quantities include the initial orbital state, the initial mantle temperatures of Io and Europa, reference silicate viscosities, melt-weakening parameters, Andrade rheological parameters, Jupiter's tidal parameters and the silicate latent heat. 
The sampled ranges are listed in Supplementary Table~1.

We set $t=0$ at the establishment of the Laplace resonance, which is assumed to have occurred early in solar system history.
Present-day analogues are therefore sought near $4.5$~Gyr, and the difference between 4.5~Gyr and the selected phase $t_\ast$ gives the inferred time of resonance establishment.
For the representative history, $t_\ast\simeq4.32$~Gyr corresponds to establishment of the resonance about $0.18$~Gyr after solar system formation.

For each simulated history, we evaluated the observational score at every sampled phase within a near-present window around $4.5$~Gyr. 
The contribution of observable $m$ at time $t$ was
\begin{equation}
\ln \mathcal{L}_m(t) = -\frac{1}{2}
\left[ \frac{x_m(t)-x_{m,\rm obs}}{\sigma_m} \right]^2.
\end{equation}
We then summed these contributions at each phase,
$\ln \mathcal{L}_{\rm joint}(t)=\sum_m \ln \mathcal{L}_m(t)$.
All observables included in the phase score were evaluated at the same phase; values from different times were not combined. 
A high score therefore favored simultaneous agreement with the full constraint set, including inward migration of Io and outward migration of Europa and Ganymede.

Because the observed state can recur at different phases of a cycle, we did not evaluate only the final timestep. 
For each simulation, the post-processing searched the near-present window ($t=4.0$--$4.5$~Gyr) and averaged the highest-scoring phase samples to define a single observational phase score. 
This reduces sensitivity to an isolated timestep and favors histories that reproduce the observed state over several phase samples.
We ranked all $\sim80{,}000$ simulations by this final score and selected the highest-scoring $1\%$ for further analysis. 
These highest-scoring models jointly reproduce the adopted astrometric and geophysical observations at a common near-present phase.
We use the combined score as a ranking statistic rather than as a formal likelihood.
All models in the highest-scoring $1\%$ develop near-present oscillatory behavior, compared with $31\%$ of the full ensemble.

\clearpage

\bibliographystylemethods{sn-nature}
\bibliographymethods{sn-bibliography}
\bmhead{Acknowledgments}

We thank D.~Stevenson (Caltech), K.~Batygin (Caltech), J.~Lunine (Caltech), M.~Efroimsky (US Naval Observatory), R.~Park (NASA JPL), and Katherine de Kleer (Caltech) for insightful discussions on the dynamical and geophysical aspects of this study that significantly improved its quality.
A.B. thanks M.~Simons (Caltech) for providing computational resources to conduct large-scale simulations.

\bmhead{Funding}
A portion of this research was carried out at the Jet Propulsion Laboratory, California Institute of Technology, under a contract with the National Aeronautics and Space Administration (80NM0018D0004).
Copyright 2026.
All rights reserved.
This work was partially funded by NASA SSW grant number NNH22ZDA001N.
Funding for S.D.V. was provided by NASA's Precursor Science Investigations for Europa (22-PSIE22\_2-0024) and Solar System Workings (23-SSW23-0080) Programs.

\subsection*{Data availability}
The data supporting the findings of this study will be deposited in a public repository upon publication. 
Data required for peer review are available from the corresponding author upon request. 

\subsection*{Code availability}
The custom code used for the coupled thermal--orbital simulations and post-processing will be deposited in a public repository upon publication. 
Code required for peer review is available from the corresponding author upon request.

\subsection*{Author contributions}
A.B. conceived and led the study, developed the coupled thermal--orbital model, numerical framework, and Europa core-formation implementation, performed the simulations and analyses, contributed to funding acquisition, and wrote the original manuscript. 
S.D.V. contributed to the study design, interpretation of the results, and manuscript preparation. 
J.F. contributed to the conceptual development of the study, developed the orbital formulation, contributed to funding acquisition and to manuscript preparation. 
B.F. contributed to the numerical simulations and manuscript preparation. M.M.D. contributed to the thermodynamic calculations underlying Europa's core-formation model. 
H.H. and G.S. developed the initial thermal--orbital model, contributed to funding acquisition, and participated in manuscript preparation. 
F.N. contributed to building the coupled thermal--orbital model, interpretation of the results, and manuscript preparation.
All authors discussed the results and reviewed and approved the final manuscript.

\subsection*{Competing interests}
The authors declare no competing interests.

\clearpage

\section*{Extended Data}

\setcounter{table}{0}

\setcounter{figure}{0}

\renewcommand{\thetable}{\arabic{table}}

\renewcommand{\thefigure}{\arabic{figure}}
\makeatletter
\renewcommand{\fnum@table}{Extended Data Table~\thetable.}

\renewcommand{\fnum@figure}{Extended Data Figure~\thefigure.}

\makeatother


\begin{table*}[ht]
\centering
\small
\renewcommand{\arraystretch}{1.15}
\begin{tabular}{p{0.34\textwidth} p{0.43\textwidth} p{0.17\textwidth}}
\hline
\textbf{Quantity} & \textbf{Adopted constraint} & \textbf{Reference} \\
\hline

Present migration rates, $\dot n/n$  & Io: $(+0.14 \pm 0.01)\times10^{-10}~\mathrm{yr^{-1}}$ \newline
Europa: $(-0.43 \pm 0.10)\times10^{-10}~\mathrm{yr^{-1}}$ \newline
Ganymede: $(-1.57 \pm 0.27)\times10^{-10}~\mathrm{yr^{-1}}$ & \cite{lainey_etal09}
\\
Laplace-resonance drift, $\dot\nu$ & $(0.74 \pm 0.24)\times10^{-7}~\mathrm{rad\,yr^{-2}}$ & \cite{lainey_etal09}
\\
Io tidal response & $\mathrm{Re}(k_2)=0.125\pm0.047$ \newline
$k_2/Q=0.0109\pm0.0054$ & \cite{park2025io}
\\
Surface heat fluxes & Io: $0.5$--$3~\mathrm{W\,m^{-2}}$ \newline
Europa: $0.015$--$0.035~\mathrm{W\,m^{-2}}$ & \cite{veeder_etal94,marchis_etal05,levin_etal26}
\\
Europa conductive ice-shell thickness & $19$--$39~\mathrm{km}$ & \cite{levin_etal26}
\\
Normalized moments of inertia, $C/(MR^2)$ & Io: $0.378\pm0.00035$ \newline Europa: $0.3547\pm0.0024$ &
\citemixed{casajus_etal21}{schubert_etal04}
\\
\\
Present mean motions & Io: $203.5~\mathrm{deg\,d^{-1}}$ \newline
Europa: $101.4~\mathrm{deg\,d^{-1}}$ \newline
Ganymede: $50.3~\mathrm{deg\,d^{-1}}$ & \cite{lainey_etal09}
\\
Present forced eccentricities & Io: $0.0041$ \newline
Europa: $0.0095$ \newline
Ganymede: $0.0008$ & \cite{lari2024nature} \\

\hline
\end{tabular}
\caption{\textbf{Observational and geophysical constraints used in the ensemble selection.}
Positive $\dot n/n$ corresponds to inward migration. 
The table lists the present-day astrometric, tidal and geophysical measurements used to compare the coupled thermal--orbital models with the Galilean system.
It should be noted that the surface heat flux of Europa proposed in \cite{levin_etal26} is based on the assumption of pure ice.}
\label{edtable:constraints}
\end{table*}



\begin{figure}
    \centering
    \includegraphics[width=0.95\linewidth]{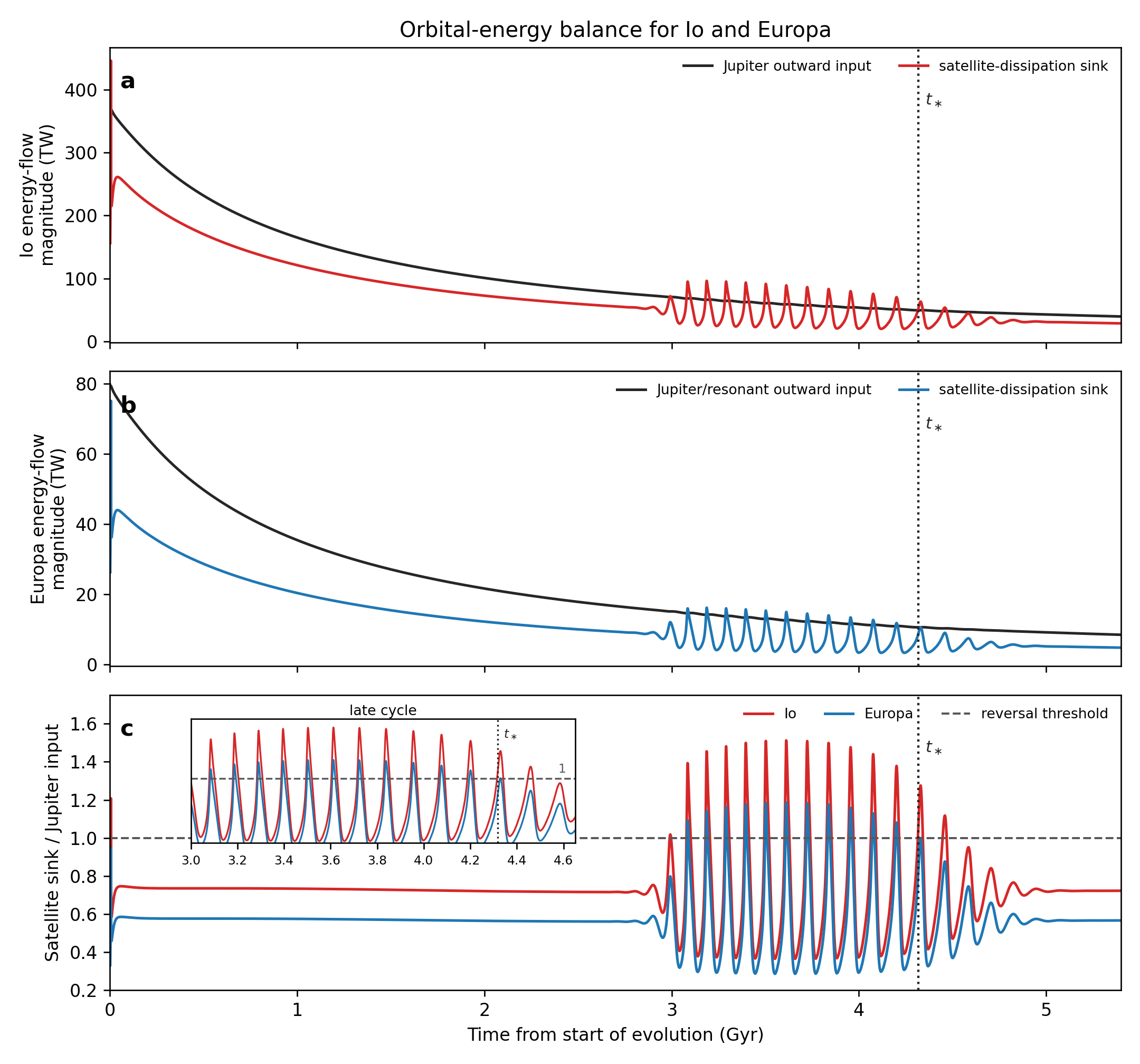}
\caption{\textbf{Orbital-energy balance controlling Io and Europa migration reversals.}
\textbf{(a)} Outward orbital-energy input associated with Jupiter's tide and inward-driving satellite-dissipation sink for Io in the representative selected history.
\textbf{(b)} Energy-flow balance for Europa.
\textbf{(c)} Ratio of the satellite-dissipation sink to the Jupiter-driven outward input for Io and Europa. 
Values above unity correspond to inward migration.
The selected phase $t_\ast$ lies after Io has crossed its reversal threshold but before Europa has done so.
The inset shows the late-cycle interval. 
Details of the energy-balance calculation are given in Supplementary Section~3.}
\label{edfig:energy_balance}
\end{figure}

\begin{figure}
    \centering
    \includegraphics[width=0.95\linewidth]{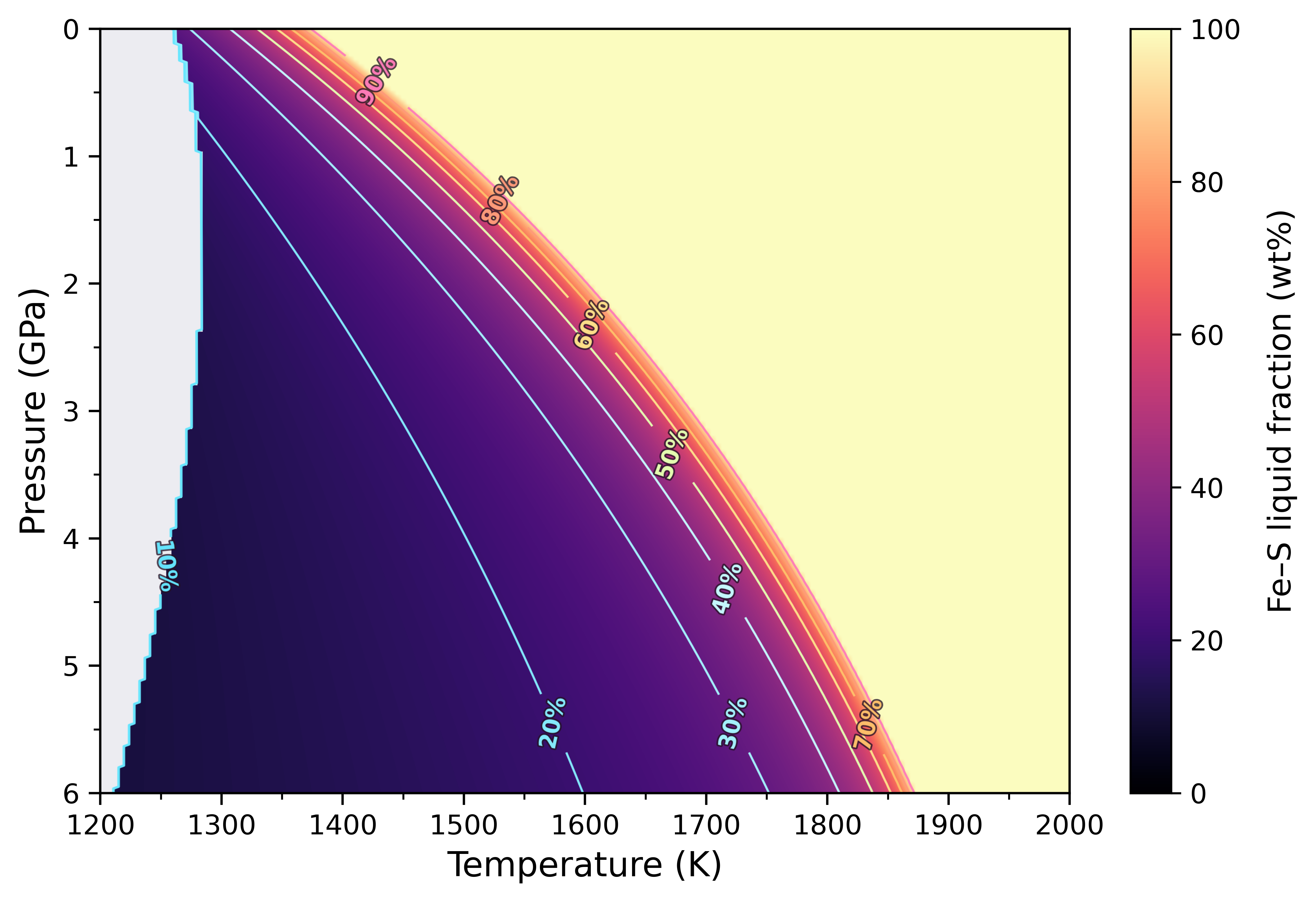}
\caption{\textbf{Fe--S thermodynamic model used for Europa core formation.}
Pressure--temperature phase diagram for the adopted CM bulk composition in the Fe--S system, showing the abundance of the Fe--S liquid phase in wt\%. 
The solidus and liquidus fits define the pressure-dependent melting interval used in the core-growth calculation. 
At each timestep, the local Fe--FeS melt fraction is computed from the layer temperature and pressure using a lever rule across this interval. 
The resulting metallic melt proxy controls the growth of Europa's segregated core and the associated differentiation heating in the coupled thermal--orbital integrations.}
\label{phase_diagram}
\end{figure}

\begin{figure}
    \centering
    \includegraphics[width=0.95\linewidth]{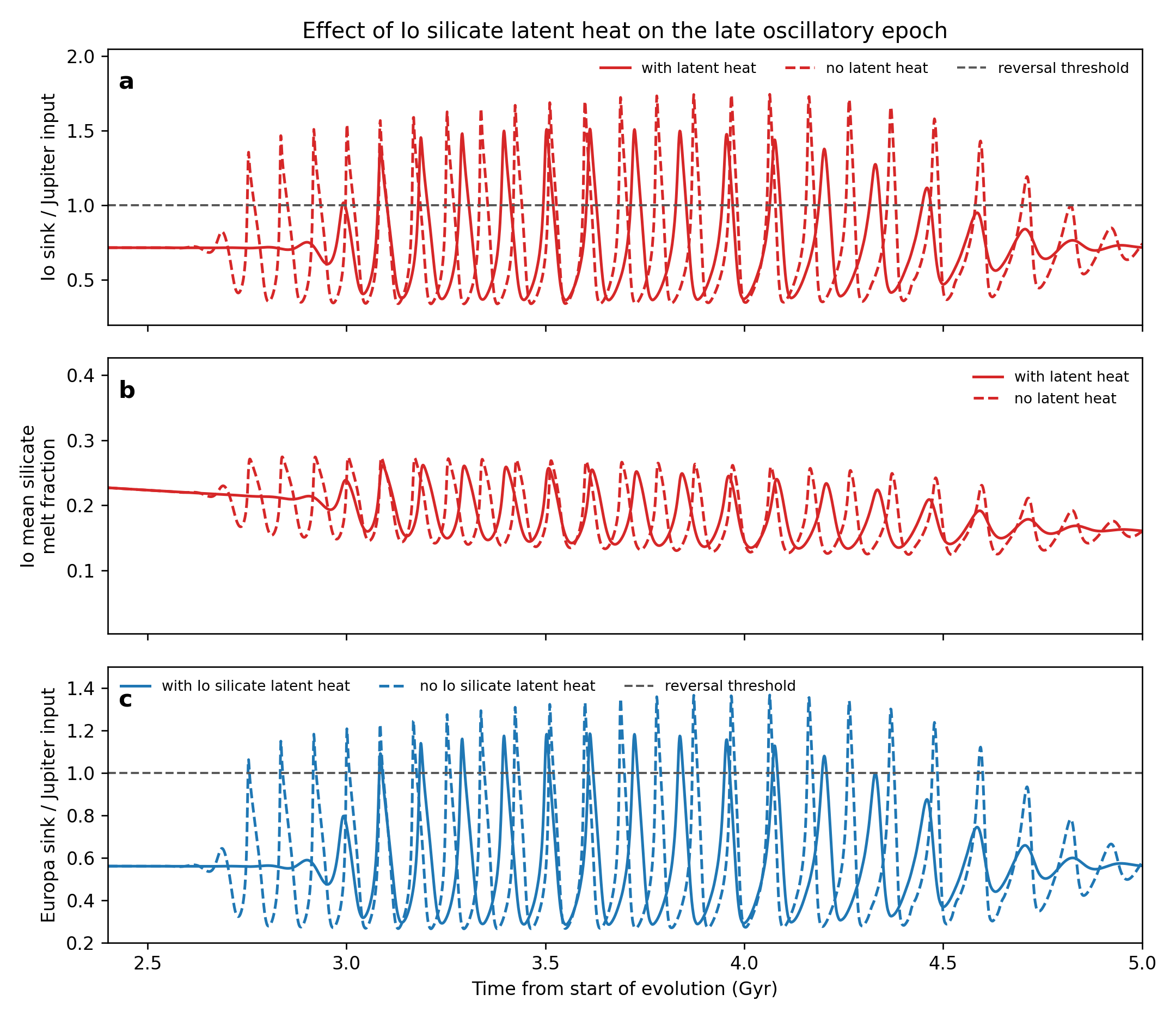}
\caption{\textbf{Effect of Io silicate latent heat on the late oscillatory epoch.}
Representative-model integrations with and without Io's silicate latent heat show that latent heat changes the timing and duration of the late limit cycle.
\textbf{(a)} Ratio of the satellite-dissipation sink to the Jupiter-driven outward orbital-energy input for Io. 
Values above unity correspond to inward migration.
\textbf{(b)} Io mean silicate melt fraction.
\textbf{(c)} Sink/input ratio for Europa, showing that the latent-heat-regulated Io cycle also shifts the timing of Europa's migration-reversal threshold crossings.}
\label{edfig:latent_ablation}
\end{figure}




\clearpage

\end{document}